\documentclass [prc,aps,showpacs]{revtex4}
\usepackage[usenames]{color}
\usepackage{ulem} 

\usepackage{graphicx}

\begin{document}

\title{Nuclear first order phase transition associated with Helmholtz free energy of canonical ensemble}

\author{A.S.~Parvan}

\affiliation{Bogoliubov Laboratory of Theoretical Physics, Joint Institute for Nuclear Research, 141980 Dubna, Russian Federation}

\affiliation{Institute of Applied Physics, Moldova Academy of Sciences, MD-2028 Chisinau, Republic of Moldova}

\begin{abstract}
It was shown that in the canonical ensemble the simple exactly soluble statistical model of nuclei decay into nucleons, which is a limiting case of the statistical multifragmentation model, predicts the nuclear first order phase transition associated with the Helmholtz free energy different from the first order phase transition of the liquid-gas type associated with the Gibbs free energy. The main thermodynamic properties of this phase transition were explored on the basis of the method of the thermodynamic potential and its first and second derivatives in the thermodynamic limit. It was established that the thermodynamic potential $F$ is a piecewise smooth function and its first order partial derivatives with respect to variables of state are piecewise continuous functions. At the points of phase transition, the energy in the caloric curve is discontinuous at the constant temperature and fixed values of the specific volume, while the pressure and the chemical potential in the equations of state are discontinuous at the constant specific volume and fixed values of the temperature.
\end{abstract}

\pacs{25.70. -z; 25.70.Mn; 05.70.Fh}

\maketitle %\narrowtext

\section{Introduction}
The heavy-ion collision experiments at RHIC, the LHC and the future FAIR and NICA projects are designed to establish the QCD phase diagram. The first order QCD phase transition, which separates the hadronic phase from the deconfined quark-gluon phase below the critical endpoint, is thought to be represented by a coexistence line on the phase diagram in the plane of temperature $T$ and baryon chemical potential $\mu$~\cite{Cleymans86,Yagi,Fukushima11,Ohnishi12,Weise12,Fischer11,Schafer05}. Such a first order phase transition, which is represented by a line in the $T-\mu$ plane, is a liquid-gas type one associated with the Gibbs free energy, which is a thermodynamic potential of the isobaric ensemble~\cite{Parvan12}. It is accepted that the nuclear multifragmentation, i.e. the disintegration of nuclei into fragments observed in intermediate-energy nuclear reactions, is also the first order phase transition of the nuclear liquid-gas type (see, e.g., ref.~\cite{Mishustin06}, and the references cited therein). However, on the theoretical side there is an ambiguity about the existence of discontinuity of energy (plateau) in the nuclear multifragmentation caloric curve in different statistical ensembles. One class of the statistical multifragmentation models (SMM) predicts a plateau in the caloric curves in the canonical and microcanonical ensembles \cite{Bondorf85,Gross90,Das2003,Parvan00}, but another class of SMM predicts a plateau in the caloric curves in the isobaric ensemble~\cite{Elliott00,Aguiar06}, which is a typical signature of the nuclear liquid-gas phase transition.
Another example of the first order phase transition of the liquid-gas type which usually happens in nuclear matter is the phase transition taking place in the interacting hadronic nuclear matter at low temperatures~\cite{Serot86,Kapusta,Parvan12}.

The first order phase transition of the liquid-gas type associated with the Gibbs free energy is defined by the cusp of the Gibbs free energy per particle (the chemical potential) and the corresponding jump discontinuities of its first derivatives, i.e. the specific volume and the entropy per particle, at constant values of pressure and temperature. Namely, the Gibbs free energy per particle is a piecewise smooth function and its first order partial derivatives with respect to variables of state of the isobaric ensemble $(T,p)$, i.e., the entropy per particle and the specific volume are the piecewise continuous functions. For the liquid-gas phase transition the energy in the caloric curve is discontinuous in the isobaric and the grand canonical ensembles at fixed values of the pressure and the chemical potential, respectively, and it is continuous in the canonical and microcanonical ensembles at fixed values of the specific volume. However, the specific volume in the isotherms is discontinuous in the isobaric and the canonical ensembles at fixed values of the temperature. The phase diagrams for the liquid-gas phase transition are represented by the coexistence lines in the planes $T-\mu$ and $T-p$ and the coexistence areas in the plane of temperature and specific volume (density of particles) and in the plane of temperature and entropy per particle.

In the Ehrenfest classification scheme the first order phase transition is defined as a phase transition which can be associated only with the Gibbs free energy of the isobaric ensemble~\cite{Ehrenfest,Stanley}. Such a phase transition is a liquid-gas type one. In refs.~\cite{Papon,Yeomans}, the Ehrenfest definition of the first order phase transition was generalized to any thermodynamic potential. In this case, it is the one associated with a finite discontinuity in one or more first derivatives of the appropriate thermodynamic potential with respect to its variables of state.
Some models which predict the first order phase transition in the canonical and microcanonical ensembles, the thermodynamical potentials of which are different from the Gibbs free energy, can be seen in refs.~\cite{Campa09,Huller94,Barre02,Gross97,Chomaz06}.

The present paper proposes a simple statistical model of total nuclei decay into nucleons, which is a limiting case of the statistical multifragmentation model which has the first order phase transition associated with the Helmholtz free energy for which the Ehrenfest definition is given in the canonical ensemble. The exact analytical results for this model were obtained. They allowed us to find the general thermodynamic properties of the first order phase transition associated with the Helmholtz free energy and determine its differences from the phase transition of the liquid-gas type. For example, the free energy per nucleon is a continuous function at the points of phase transition, but the first order partial derivatives of the thermodynamic potential of the canonical ensemble with respect to variables of state, i.e. the entropy per nucleon, the pressure and the chemical potential, have jump discontinuities.

The structure of the paper is as follows. In Section II, we briefly describe basic ingredients of the statistical model of total nuclei decay into nucleons. The thermodynamic results for the first order phase transition associated with the Helmholtz free energy and the caloric curve are discussed in Sections III. The main conclusions are summarized in the final section.

\section{The statistical model of the nuclei decay into nucleons}\label{Nucl}
\subsection{The general formalism}
The partition function of the statistical multifragmentation model in the canonical ensemble which describes the decay of one nucleus of $A$ nucleons into different nuclear fragments in the volume $V$ at the temperature $T$ is actually given by the expression~\cite{Parvan00,DasGupta98,Das05,Parvan99,Parvan04}
\begin{equation}\label{1a}
   Z_{A} = \sum\limits_{\{n_{k}\}} \delta\left(\sum\limits_{k=1}^{A} k n_{k} - A\right) \ \prod_{k=1}^{A} \frac{\omega_{k}^{n_{k}}}{n_{k}!} =
  \frac{1}{A} \ \sum\limits_{k=1}^{A} k \omega_{k} Z_{A-k}
\end{equation}
and
\begin{equation}\label{2a}
  \omega_{k} = g_{k} V_{f} \left(\frac{m k T}{2\pi}\right)^{3/2} \ e^{\frac{W_{k}}{T}},
\end{equation}
where $Z_{0}=1$, $\omega_{k}$ is the partition function of a fragment (nucleus) which has $k$ nucleons, $V_{f}=V-v_{0}A$ is the free volume, $W_{k}$ is the binding energy of the fragment of $k$ nucleons, $m_{N}$ is the nucleon mass, $v_{0}=1/\rho_{0}$, $\rho_{0}$ is the normal nuclear density and $g_{k}$ is the spin-isospin degeneracy factor. The binding energy of the free nucleons $W_{1}=0$. Note that throughout the paper we use the system of natural units, $\hbar=c=k_{B}=1$.

The partition function (\ref{1a}) in the limiting case of the SMM, when one nucleus of $A$ nucleons decays into free nucleons only, $\omega_{2}=\ldots =\omega_{A-1}=0$, can be written as
\begin{equation}\label{3a}
 Z_{A} = \frac{\omega_{1}^{A}}{A!} +\frac{\omega_{A}^{1}}{1!}.
\end{equation}

Let us generalize this limiting case of the SMM to the system of $N$ number of nuclei of $A$ nucleons which decay into $B$ free nucleons, $B=NA$, in a volume $V$, in contact with a heat reservoir of temperature $T$. Considering the definition (\ref{3a}) and the Maxwell-Boltzmann statistics of nuclei and nucleons, we can write the thermodynamical potential, the Helmholtz free energy, and the partition function in the canonical ensemble $(T,V,B)$ as
\begin{equation}\label{1}
    F = -T \ln Z_{B}
\end{equation}
and
\begin{equation}\label{2}
    Z_{B} = \frac{\omega_{1}^{B}}{B!} + \frac{\omega_{A}^{N}}{N!}.
\end{equation}
Here, the free volume $V_{f}=V-v_{0}B$ is used. The mean number of the free nucleons $\langle n_{1}\rangle$ and the mean number of the initial free nuclei $\langle n_{A} \rangle$ are given by
\begin{equation}\label{4}
\langle n_{1}\rangle = \frac{B}{Z_{B}} \frac{\omega_{1}^{B}}{B!}, \qquad
\langle n_{A} \rangle = \frac{N}{Z_{B}} \frac{\omega_{A}^{N}}{N!}.
\end{equation}
The total baryon charge of the system $\langle B\rangle$ and the mean multiplicities of particles in the system $\langle m\rangle$ are
\begin{eqnarray}\label{5}
B &=& \langle n_{1}\rangle + A\langle n_{A}\rangle, \\ \label{6}
 \langle m\rangle &=& \langle n_{1}\rangle+\langle n_{A}\rangle.
\end{eqnarray}

The first order partial derivatives of the thermodynamic potential $F$ with respect to variables of state of the canonical ensemble $(T,V,B)$, i.e. the entropy $S$, the pressure $p$ and the chemical potential $\mu$ can be written as
\begin{eqnarray}\label{7}
  S &=& -\left(\frac{\partial F}{\partial T}\right)_{VB}= \frac{E-F}{T}, \\    \label{8}
  p &=& -\left(\frac{\partial F}{\partial V}\right)_{TB}= \frac{T}{V_{f}} \ \langle m \rangle, \\    \label{9}
  \mu &=&  \left(\frac{\partial F}{\partial B}\right)_{TV}= p v_{0} + \frac{T}{B} \left\{1+ \left[-\ln\omega_{1}+\psi(B)\right] \langle n_{1} \rangle + \left[-\ln\omega_{A}+\psi(N)\right] \langle n_{A} \rangle \right\},
\end{eqnarray}
where $\psi(z)$ is the psi-function. The mean energy of the system is given by
\begin{equation}\label{10}
  E =  -T^{2} \left(\frac{\partial}{\partial T}\frac{F}{T}\right)_{VB} = \frac{3}{2} T \ \langle m \rangle -  \langle n_{A} \rangle \ W_{A}.
\end{equation}
The second order partial derivatives of the thermodynamic potential $F$ with respect to variables of state, i.e. the heat capacity $C_{V}$, the compressibility $k_{T}$ and the susceptibility $\chi$,  can be written as
\begin{eqnarray}\label{11}
  C_{V} &=& -T \left(\frac{\partial^{2}F}{\partial T^{2}}\right)_{VB}= \frac{3}{2} \ \langle m \rangle + \langle n_{1}\rangle\langle n_{A}\rangle \
\frac{1}{A}\left[\frac{3}{2}(A-1)+\frac{W_{A}}{T}\right]^{2}, \\ \label{12}
\frac{1}{k_{T}V} &=& \left(\frac{\partial^{2}F}{\partial V^{2}}\right)_{TB}= \frac{p}{V_{f}} - \langle n_{1}\rangle\langle n_{A}\rangle \
\frac{T}{V_{f}^{2}}\frac{(A-1)^{2}}{A}, \\ \label{13}
\frac{1}{\chi} &=& \left(\frac{\partial^{2}F}{\partial B^{2}}\right)_{TV}= \frac{T}{B^{2}} \left\{ \frac{v_{0}B}{V_{f}} \left(2+\frac{v_{0}B}{V_{f}}\right)\langle m\rangle -1 +  B\frac{\partial \psi(B)}{\partial B}\langle n_{1}\rangle + N\frac{\partial \psi(N)}{\partial N}  \langle n_{A}\rangle  \right\} \nonumber \\
  &-& \langle n_{1}\rangle\langle n_{A}\rangle  \frac{T}{B^{2}} A \left[-\frac{A-1}{A}\frac{v_{0}B}{V_{f}}+\ln\omega_{1}-\psi(B)-\frac{1}{A}(\ln\omega_{A}-
  \psi(N)) \right]^{2}.
\end{eqnarray}
The second derivatives allow us to identify the convexity properties of the thermodynamic potential $F$ which are related to the thermal, mechanical and particle stability of the system~\cite{Stanley}.

\subsection{The thermodynamic limit}
Let us consider the thermodynamic limit: $N\to\infty$, $V\to\infty$ and $v=V/B=const$. Then the free energy per nucleon (\ref{1}) reduces to
\begin{eqnarray}\label{14}
  f &=& \frac{F}{B}=-\frac{T}{A}\ln\left(z_{1}^{N}+z_{A}^{N}\right)^{\frac{1}{N}}, \quad N\to\infty,  \\ \label{15}
  z_{1} &=& \left[g_{1}v_{f}e \left(\frac{mT}{2\pi}\right)^{3/2}\right]^{A},  \\ \label{16}
  z_{A} &=& g_{A}v_{f}e \left(\frac{mT}{2\pi}\right)^{3/2} A^{5/2} e^{\frac{W_{A}}{T}},
\end{eqnarray}
where $v_{f}=V_{f}/B=v-v_{0}$. The mean number of the free nucleons and the mean number of the free nuclei (\ref{4}) then become
\begin{equation}\label{17}
\langle n_{1}\rangle = B \frac{1}{1+e^{-aN}}, \qquad
\langle n_{A} \rangle = N \frac{1}{1+e^{aN}},
\end{equation}
where
\begin{equation}\label{18}
    a=\ln\frac{z_{1}}{z_{A}}.
\end{equation}
The first order partial derivatives of the Helmholtz potential $F$ with respect to variables of state, i.e. the entropy per nucleon (\ref{7}), the pressure (\ref{8}) and the chemical potential (\ref{9}), in the thermodynamic limit take the form
\begin{eqnarray}\label{19}
s &=&  \frac{S}{B}= \frac{\varepsilon-f}{T}, \\ \label{20}
  p &=& \frac{T}{v_{f}} \ \frac{\langle m\rangle}{B},  \\ \label{21}
  \mu &=& p v- \frac{T}{A} \left[ \ln z_{1} \frac{\langle n_{1}\rangle}{B}+ \ln z_{A} \frac{\langle n_{A}\rangle}{N}\right],
\end{eqnarray}
The mean energy per nucleon (\ref{10}) in the thermodynamic limit becomes
\begin{equation}\label{22}
   \varepsilon = \frac{E}{B} = \frac{3}{2} T \frac{\langle m\rangle}{B} - \frac{\langle n_{A}\rangle}{N} \frac{W_{A}}{A}.
\end{equation}
Finally, the second order partial derivatives of the thermodynamic potential $F$ with respect to variables of state, i.e. the heat capacity per nucleon (\ref{11}), the compressibility (\ref{12}) and the susceptibility (\ref{13}), in the thermodynamic limit can be written as
\begin{eqnarray}\label{23}
 c_{v} &=& \frac{C_{V}}{B} = \frac{3}{2} \frac{\langle m\rangle}{B} + \frac{\langle n_{1}\rangle}{B} \frac{\langle n_{A}\rangle}{N} \frac{B}{A^{2}} \left[
 \frac{3}{2}(A-1) + \frac{W_{A}}{T}\right]^{2}, \\ \label{24}
  \frac{1}{k_{T}v} &=&  \frac{p}{v_{f}} -
  \frac{\langle n_{1}\rangle}{B} \frac{\langle n_{A}\rangle}{N} \frac{TB}{v_{f}^{2}} \left(\frac{A-1}{A}\right)^{2},  \\ \label{25}
  \frac{B}{\chi} &=& \frac{p v^{2}}{v_{f}} - \frac{\langle n_{1}\rangle}{B} \frac{\langle n_{A}\rangle}{N} \frac{TB}{A^{2}} \left[a -
  (A-1)\frac{v}{v_{f}}\right]^{2}.
\end{eqnarray}
Note that in this model the thermodynamic limit is taken in accordance with the standard rules of the thermodynamics and the statistical mechanics~\cite{Huang}.

\section{The first order phase transition in the canonical ensemble}\label{Rez1}
Let us describe the phase transition of the statistical model of the nuclei decay into nucleons in the canonical ensemble. This ensemble is characterized by the potential, $F(T,V,B)$ and the physical observables correspond to its first and second order partial derivatives. The geometry (curvature) of the hypersurface $F$ determines the physical properties of the system. In the thermodynamic limit at three different values of the variable $a$, $a<0$, $a=0$ and $a>0$, the thermodynamic potential per nucleon (\ref{14}) is a homogeneous function of the first degree of the extensive variable of state $B$:
\begin{equation}\label{26}
         F(T,V,B) =  B f(T,v).
\end{equation}
Then in the thermodynamic limit at three different values of the variable $a$ the first and the second order partial derivatives of the thermodynamic potential $F$ with respect to variables of state $(T,V,B)$ (\ref{17})--(\ref{23}) can be written as
\begin{eqnarray}\label{27}
    s &=& -f'_{T}(T,v), \qquad \;\;\;\;\;\;\; p= -f'_{v}(T,v), \qquad \quad \mu= f+pv, \\ \label{28}
    \frac{c_{v}}{T} &=& -f''_{TT}(T,v), \qquad  \frac{1}{k_{T}v} = f''_{vv}(T,v), \qquad   \frac{B}{\chi v^{2}} = f''_{vv}(T,v),
\end{eqnarray}
where $f'_{x}(x,y)=(\partial f/\partial x)_{y}$ and $f''_{xx}(x,y)=(\partial^{2} f/\partial x^{2})_{y}$. These quantities satisfy the differential equation for $f$, the fundamental equation of thermodynamics and the Euler theorem
\begin{equation}\label{29}
  df = -sdT-pdv,  \qquad Tds = d\varepsilon + p dv, \qquad  Ts=\varepsilon+pv-\mu.
\end{equation}
Moreover, the Legendre transform, $f=\varepsilon-Ts$, and the relation for second derivatives, $B/\chi=v/k_{T}$, are valid. Note that eqs.~(\ref{26})--(\ref{29}) are satisfied separately in the regions $a<0$, $a=0$ and $a>0$.

The values of $a<0$ give the homogeneous phase $I$ of the system which consists of the free nuclei $\langle n_{1}\rangle/B=0$ and $\langle n_{A}\rangle/N=1$. The homogeneous phase $II$ of the system is determined by the values of $a>0$ in which the system is composed by the free nucleons $\langle n_{1}\rangle/B=1$ and $\langle n_{A}\rangle/N=0$. The values of $a=0$ in Eq.~(\ref{18}) determine the line of phase boundary in the $T-v$ space
\begin{equation}\label{30}
    v= v_{0} + \left(\frac{g_{1}^{A}}{g_{A}}A^{-5/2}\right)^{-\frac{1}{A-1}}\left(\frac{mT}{2\pi}\right)^{-\frac{3}{2}} \ e^{\frac{1}{A-1}\frac{W_{A}}{T}-1}.
\end{equation}

\begin{figure}[htp]
\includegraphics[width=14cm]{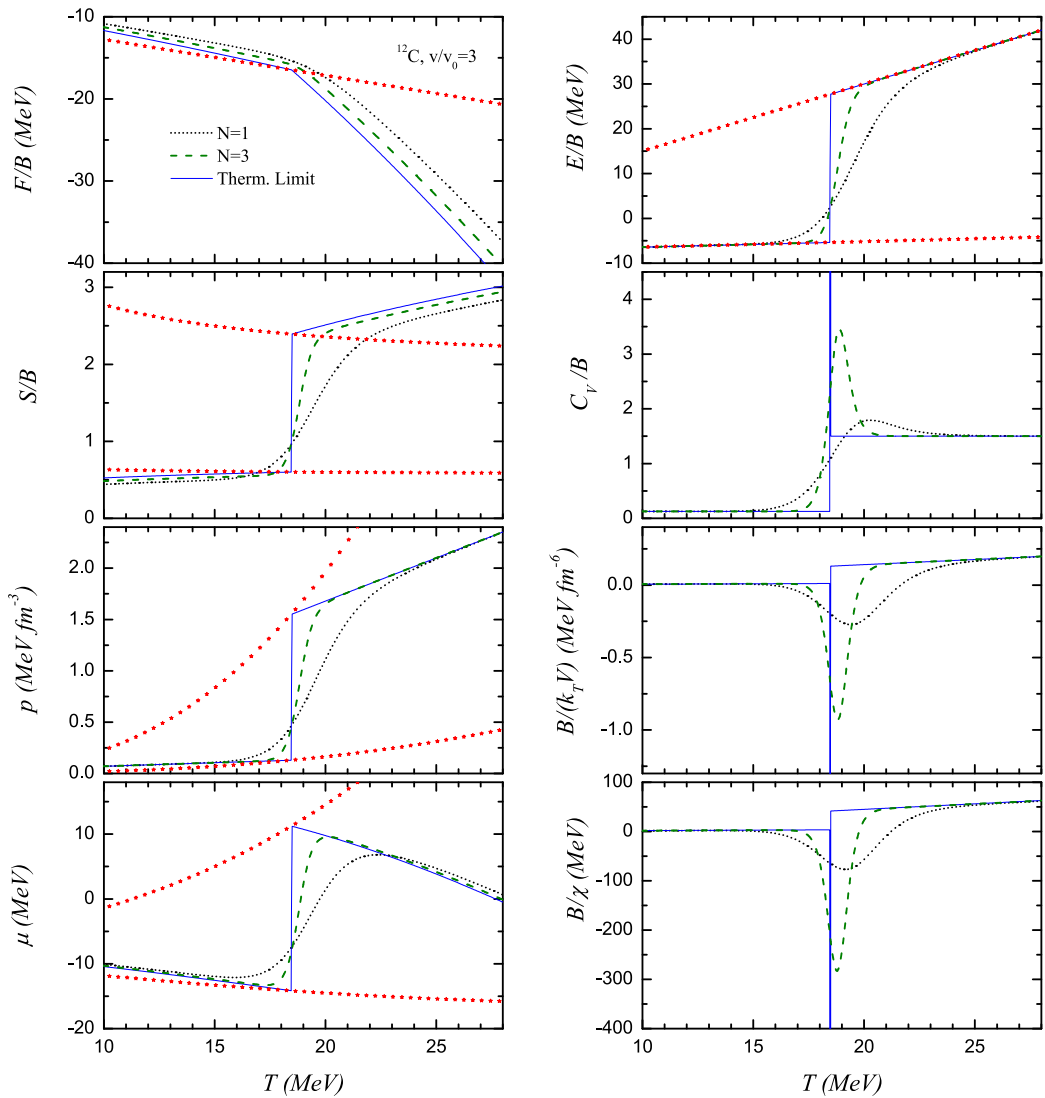} \vspace{-0.3cm}
\caption{(Color online) The thermodynamic potential per nucleon $f$, the entropy per nucleon $s$, the pressure $p$, the chemical potential $\mu$, the energy per nucleon $\varepsilon$, the heat capacity per nucleon $c_{v}$, the compressibility $1/k_{T}v$ and the susceptibility per nucleon $B/\chi$ as functions of the temperature $T$ at fixed specific volume $v$ for the finite number $N$ of nuclei and the thermodynamic limit $N=\infty$. The dotted, dashed and solid curves were calculated by the statistical model of nuclei decay into nucleons in the canonical ensemble for the $^{12}C$ nuclei with $N=1,3$ and $N=\infty$, respectively, at the specific volume $v=3v_{0}$. The symbols are the phase diagrams in the thermodynamic limit.} \label{f1}
\end{figure}
Figure~\ref{f1} presents the behavior of the free energy per nucleon $f$ and its first and second partial derivatives with respect to variables of state as functions of the temperature $T$ at fixed specific volume $v$ for the $^{12}C$ nuclei which decay into nucleons for the finite number $N$ of nuclei and in the thermodynamic limit, $N=\infty$. In the calculations we shall assume the typical numerical values of the spin-isospin degeneracy factors adopted in the SMM~\cite{Bondorf95}, i.e. $g_{1}=4$ and $g_{A}=1$. Therewith, the binding energy per nucleon for the nucleus of $^{12}C$ is taken to be $W_{A}/A=7.680144$ MeV. The function $f(T)$ is continuous for all $T$ and has a cusp at the point of phase transition $T=T^{*}$ where the first order partial derivatives, i.e. the entropy per nucleon $s(T)$, the pressure $p(T)$ and the chemical potential $\mu(T)$, and the energy per nucleon $\varepsilon(T)$ have jump discontinuities. At this point $T^{*}$ the second order partial derivatives, i.e. the heat capacity per nucleon $c_{v}(T)$, the compressibility $1/k_{T}(T)v$ and the susceptibility per nucleon $B/\chi(T)$, have the local extremum (infinite maximum or minimum). Therefore, in conformity with the Ehrenfest definition of the phase transitions the point $T=T^{*}$ is the point of the first order phase transition related to the thermodynamic potential $F$ of the canonical ensemble, i.e. the Helmholtz free energy. The jump of the entropy per nucleon at the point of phase transition $T^{*}$ is related to the latent heat. In this nuclear system two homogeneous phases can be distinguished. In the homogeneous phase $(I)$ at the temperature $T<T^{*}$ the system consists of the $N$ free nuclei of $^{12}C$, $\langle n_{A}\rangle=N$. However, in the homogeneous phase $(II)$ at the temperature $T>T^{*}$ the system is composed of the free nucleons only $\langle n_{1}\rangle=B$. At $T=T^{*}$ the mixed phase is realized. Note that in the canonical ensemble at fixed values of the volume and the number of nuclei the pressure of the system is discontinuous at the points of phase transition due to the jump increase in the multiplicity of particles which carry the partial pressure $\sim T/V_{f}$. The discontinuity of the chemical potential is also related to the internal increase in the number of free particles.

\begin{figure}[htp]
\includegraphics[width=14cm]{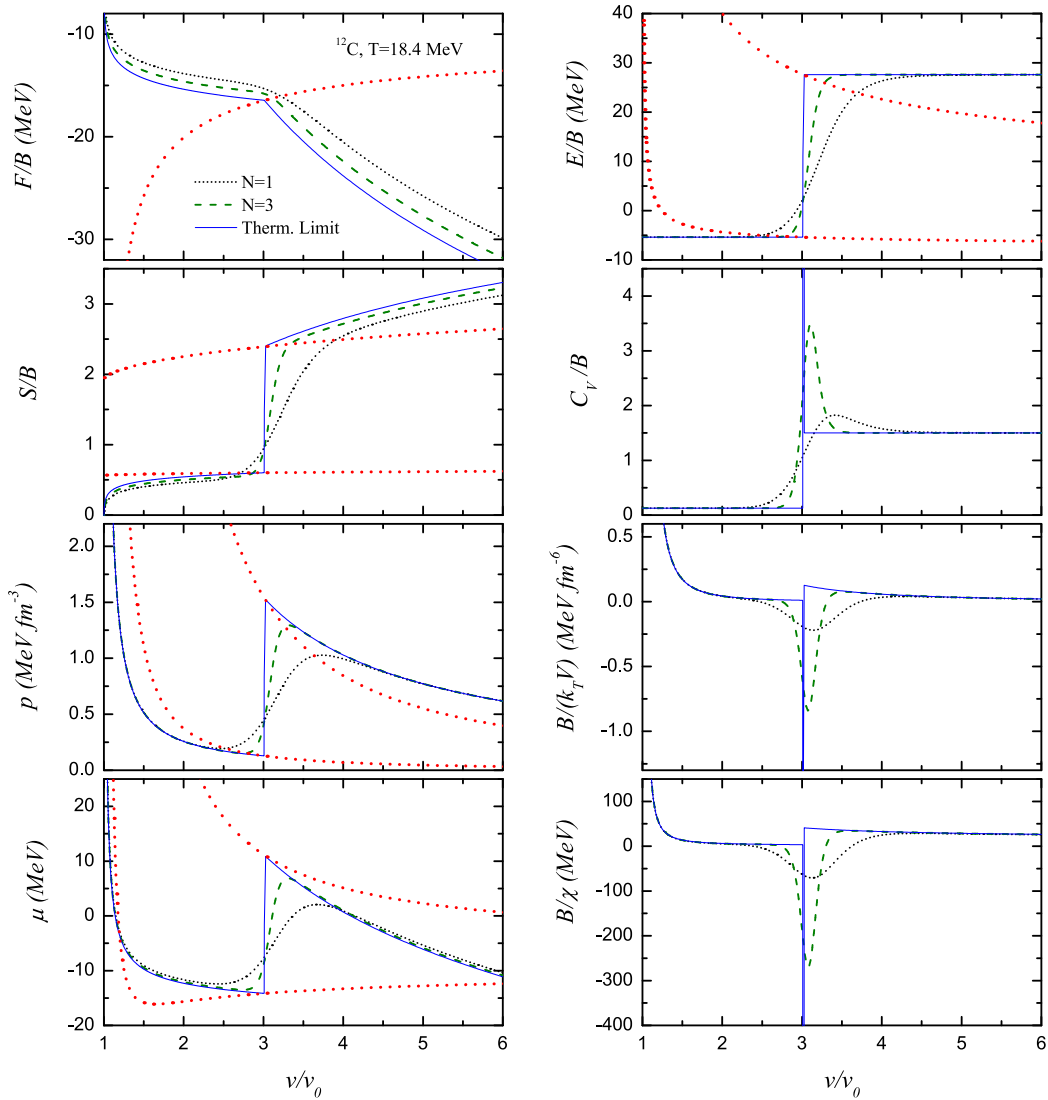}  \vspace{-0.3cm}
\caption{(Color online) The thermodynamic potential per nucleon $f$, the entropy per nucleon $s$, the pressure $p$, the chemical potential $\mu$, the energy per nucleon $\varepsilon$, the heat capacity per nucleon $c_{v}$, the compressibility $1/k_{T}v$ and the susceptibility per nucleon $B/\chi$ as functions of the specific volume $v$ at fixed temperature $T$ for both the finite number $N$ of nuclei and the thermodynamic limit, $N=\infty$. The dotted, dashed and solid curves were calculated by the statistical model of nuclei decay into nucleons in the canonical ensemble for the $^{12}C$ nuclei with $N=1,3$ and $N=\infty$, respectively, at the temperature $T=18.4$ MeV. The symbols are the phase diagrams in the thermodynamic limit.} \label{f2}
\end{figure}
Figure~\ref{f2} presents the behavior of the free energy per nucleon $f$ and its first and second partial derivatives with respect to variables of state as functions of the specific volume $v$ at fixed temperature $T$ for the $^{12}C$ nuclei which decay into nucleons for the finite number $N$ of nuclei and in the thermodynamic limit, $N=\infty$. In the thermodynamic limit the function $f(v)$ is a piecewise smooth function on a closed  interval $I$ around the point of phase transition $v=v^{*}$, i.e., it is a one-valued continuous broken-line function with the point of discontinuity of the first derivative at $v=v^{*}$. The graph has a cusp and there is no derivative. The first order partial derivatives of the thermodynamic potential per nucleon with respect to variables of state, the entropy per nucleon $s(v)$, the pressure $p(v)$ and the chemical potential $\mu(v)$ are single-valued piecewise continuous functions for all $v$. At the point $v=v^{*}$ the first order partial derivatives, i.e. the entropy per nucleon $s(v)$, the pressure $p(v)$ and the chemical potential $\mu(v)$, and the energy per nucleon $\varepsilon(v)$ have jump discontinuities. However, the second order partial derivatives of the thermodynamic potential per nucleon with respect to variables of state, i.e. the heat capacity per nucleon $c_{v}(v)$, the compressibility $1/k_{T}(v)v$ and the susceptibility per nucleon $B/\chi(v)$, have the local extremum (infinite maximum or minimum). Therefore, the point of discontinuity of the first derivatives at $v=v^{*}$ is the point of the first order phase transition related to the thermodynamic potential $F$ of the canonical ensemble, i.e. the Helmholtz free energy. The jump of the entropy per nucleon at the point of phase transition $v^{*}$ is related to the latent heat. In the nuclear system described by the statistical model of the nuclei decay into nucleons in the canonical ensemble two homogeneous phases are realized. In the homogeneous phase $(I)$ at the specific volume $v<v^{*}$ the system consists of the $N$ free nuclei, $\langle n_{A}\rangle=N$. However, in the homogeneous phase $(II)$ at the specific volume $v>v^{*}$ the system is composed of the free nucleons, $\langle n_{1}\rangle=B$. At $v=v^{*}$ the mixed phase is realized.

\begin{figure}
\includegraphics[width=8cm]{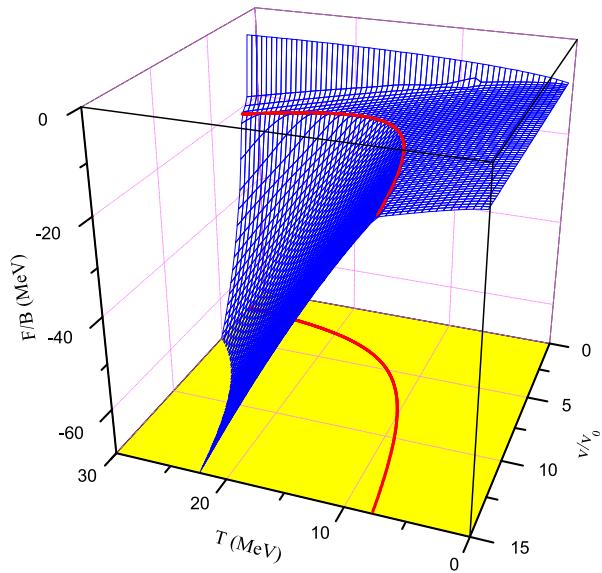} \vspace{-0.3cm}
\caption{(Color online) The Helmholtz free energy per nucleon $f(T,v)$ as a function of the temperature $T$ and the specific volume $v$ for the statistical model of nuclei decay into nucleons in the canonical ensemble in the thermodynamic limit for the $^{12}C$ nuclei. The solid lines are the line of phase transition on the surface $f(T,v)$ and its projection onto the plane $T-v$, i.e. the phase diagram.} \label{f3}
\end{figure}
Figure~\ref{f3} shows a picture of the surface of the thermodynamic potential per nucleon $f$ as a function of the variables of state $(T,v)$ near the line of phase transition which is determined by eq.~(\ref{30}). This line is a line of fracture of the surface $f$ as in all its points the function $f(T,v)$ is continuous but undifferentiated. It defines the first order phase transition associated with the Helmholtz free energy, and its projection onto the $T-v$ plane gives the corresponding phase diagram.

\begin{figure}
\includegraphics[width=12cm]{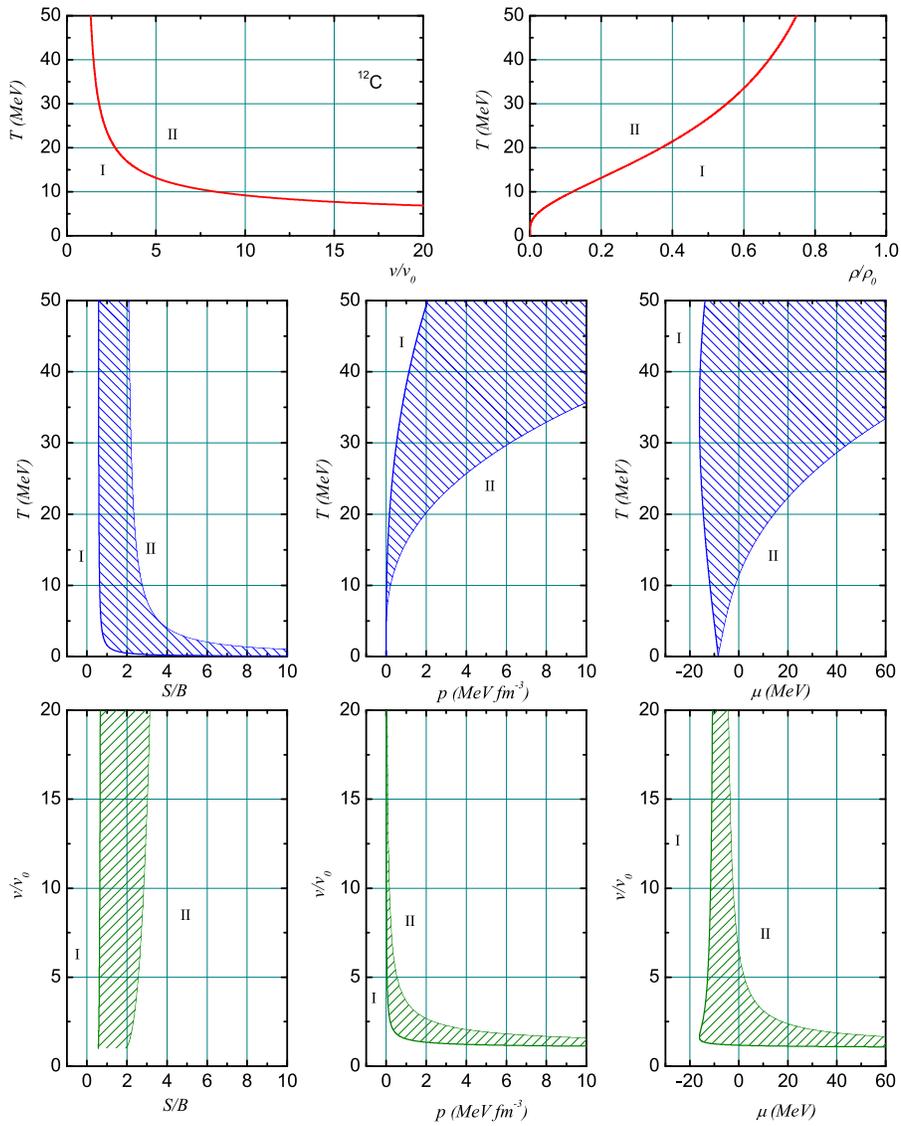} \vspace{-0.3cm}
\caption{(Color online) The phase diagrams $T-v$, $T-\rho$ (top panels), $T-s$, $T-p$, $T-\mu$ (mean panels) and $v-s$, $v-p$, $v-\mu$ (bottom panels) for the first order phase transition of the statistical model of the nuclei decay into nucleons for the $^{12}_{6}C$ nuclei in the canonical ensemble. Roman numerals denote the homogeneous phases ($I$) and ($II$), the shaded areas correspond to the mixed phase and the lines are the coexistence curves.} \label{f4}
\end{figure}
More concretely, the phase diagrams for the first order phase transition of the statistical model of the nuclei decay into nucleons in the canonical ensemble $(T,v)$ in the thermodynamic limit are depicted in Fig.~\ref{f4}. The phase diagram $T-v$ for the first order phase transition in the canonical ensemble is represented by the continuous coexistence curve which begins at the critical point $(T_{c}=\infty,v_{c}=v_{0})$ and finishes at the point $(T=0,v=\infty)$. The coexistence curve $T-\rho$ crosses the $\rho$ axes at the point $(T=0$, $\rho=0)$ and finishes at the critical point $(T_{c}=\infty,\rho_{c}=\rho_{0})$. Along these coexistence curves, the functions $s(T,v)$, $p(T,v)$ and $\mu(T,v)$, are discontinuous. Therefore, in the phase diagrams $T-s$, $T-p$, $T-\mu$, $v-s$, $v-p$ and $v-\mu$ we have the coexistence regions instead of the coexistence lines because the variables $s$, $p$ and $\mu$ are undefined at the points of phase transition $(T^{*},v^{*})$. We have two homogeneous phases in the system: the phase $(I)$ of the free nuclei and  the phase $(II)$ of the free nucleons. The mixed phase is defined by the coexistence lines on the phase diagrams $T-v$, $T-\rho$ and the coexistence areas on the phase diagrams $T-s$, $T-p$ and $T-\mu$. In contrast to this phase transition, the first order phase transition of the liquid-gas type associated with the Gibbs free energy $G$ is described by the coexistence lines on the phase diagrams $T-\mu$ and $T-p$ and the coexistence areas on the phase diagrams $T-v$, $T-\rho$~\cite{Parvan12}. From the phase diagrams shown in Fig.~\ref{f3}, the Gibbs phase rule for the first order phase transition associated with the Helmholtz free energy $F$ of the canonical ensemble can be defined by
\begin{equation}\label{31}
    T_{I}=T_{II}, \qquad v_{I}=v_{II},
\end{equation}
where the roman numerals ($I$) and ($II$) denote the homogeneous phases in equilibrium. It is characterized by the equality of temperature and specific volume across phase boundaries. In contrast to this, the Gibbs phase rule for the first order phase transition of the liquid-gas type associated with the Gibbs free energy is represented by the equality of temperature, pressure and chemical potential across phase boundaries~\cite{Landau}.

\begin{figure}
\includegraphics[width=13.5cm]{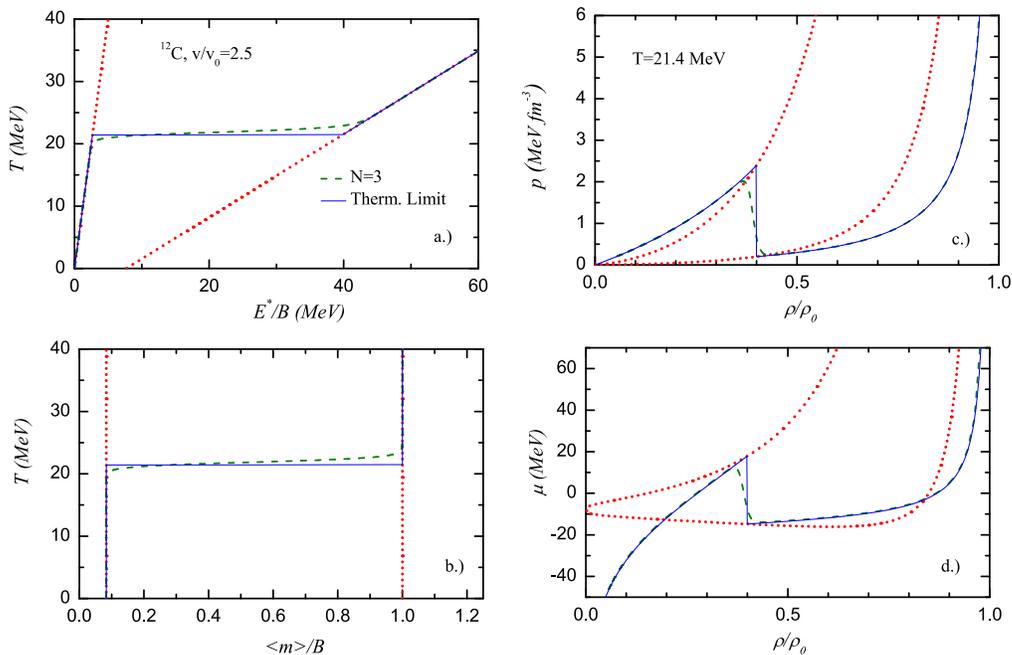} \vspace{-0.3cm}
\caption{(Color online) (a) The temperature $T$ as a function of the excitation energy per nucleon $\varepsilon^{*}$ (the caloric curve) and (b) the temperature $T$ as a function of the multiplicity $\langle m\rangle$ for the statistical model of the nuclei decay into nucleons for the $^{12}_{6}C$ nuclei in the canonical ensemble for $N=3$ and in the thermodynamic limit at the specific volume $v=2.5 v_{0}$. (c) The isotherms and (d) the dependence of the chemical potential $\mu$ on the baryon density $\rho$ in the canonical ensemble for $N=3$ and in the thermodynamic limit at the temperature $T=21.4$ MeV.}   \label{f5}
\end{figure}
Figure~\ref{f5} presents the caloric curve, the temperature $T$ as a function of the multiplicity, the equation of state or the isotherms, i.e. the dependence $p-\rho$ at fixed temperature $T$, and the dependence $\mu-\rho$ in the canonical ensemble for the first order phase transition of the statistical model of the nuclei decay into nucleons. The excitation energy per nucleon is defined by the equation, $\varepsilon^{*}=\varepsilon(T)-\varepsilon(0)$, where $\varepsilon(0)=-W_{A}/A$ is the energy per nucleon in the ground state. At the point of phase transition at temperature $T=T^{*}$ the excitation energy per nucleon $\varepsilon^{*}$ in the caloric curve has a jump discontinuity in the canonical ensemble at fixed specific volume $v$ or baryon density $\rho$. The curve $\varepsilon^{*}(T)$ is a piecewise continuous function. The multiplicity of particles in Fig.~\ref{f5}(b) has a jump discontinuity at the point of phase transition. In the phase (I) it is equal to $\langle m\rangle=N$, but in the phase (II) the multiplicity is $\langle m\rangle=B$. The pressure $p$ and the chemical potential $\mu$ in the isotherms and the equation $\mu-\rho$, respectively, are discontinuous in the canonical ensemble at the points of phase transition. See Fig.~\ref{f5}(c) and (d). Note that in the case of the first order phase transition of the liquid-gas type in the canonical ensemble, the excitation energy per nucleon $\varepsilon^{*}$ in the caloric curve is continuous at fixed values of the specific volume and the specific volume $v$ in the isotherms is discontinuous at constant values of the pressure $p$ and the chemical potential $\mu$~\cite{Parvan12}. In the canonical ensemble the liquid-gas phase transition is defined by the plateaus in the isotherms and the equation $\mu-\rho$ at fixed temperature $T$.

Summarizing, we have found that the phase transition of the statistical model of the nuclei decay into nucleons in the canonical ensemble is the first order phase transition defined by the Helmholtz free energy, instead of the Gibbs free energy which defines the first order phase transition of the liquid-gas type.  The first order phase transition associated with the Helmholtz free energy is defined by the piecewise smooth function of the Helmholtz potential per nucleon $f$ and the piecewise continuous functions of the first order partial derivatives of $F$ with respect to variables of state, the entropy per nucleon $s$, the pressure $p$ and the chemical potential $\mu$. At the points of phase transition given by the variables of state $(T,v)$ the potential $f$ is a continuous function which has a cusp, however, the first derivatives $s$, $p$ and $\mu$ (order parameters) have jump discontinuities and the second order partial derivatives $c_{v}$, $1/k_{T}v$ and $B/\chi$ have an infinite maximum or minimum. The excitation energy per nucleon $\varepsilon^{*}$ in the caloric curve is discontinuous in the canonical ensemble.

\section{Conclusions}\label{Concl}
To conclude, in this paper the statistical model of the nuclei decay into nucleons in the canonical ensemble as the limiting case of the statistical multifragmentation model was introduced. The exact analytical results for both the finite system of $N$ nuclei and the infinite system in the thermodynamic limit were obtained. It was revealed that the statistical model of the nuclei decay into nucleons in the canonical ensemble in the thermodynamic limit has the first order phase transition associated with the Helmholtz free energy instead of the Gibbs free energy. This phase transition was investigated on the basis of the method of the thermodynamic potentials and their first and second derivatives. The main thermodynamic properties of this first order phase transition in the canonical ensemble were found. These properties are as follows. The Helmholtz free energy per nucleon is the piecewise smooth function and its first order partial derivatives with respect to variables of state, i.e., the entropy per nucleon, the pressure and the chemical potential, are the piecewise continuous functions. At the points of phase transition, the free energy per nucleon is a continuous function which has a cusp both as a function of $v$ at fixed $T$ and as a function of $T$ at fixed $v$; however, the first order partial derivatives of the free energy with respect to variables of state, i.e. the entropy per nucleon, the pressure and the chemical potential, have jump discontinuities. The second order partial derivatives of the thermodynamical potential, i.e. the heat capacity per nucleon, the compressibility and the susceptibility per nucleon, have the infinite maximum or minimum at the points of phase transition. The phase diagrams $T-v$ and $T-\rho$ are represented by the coexistence lines; however, the phase diagrams $T-s$, $T-p$, $T-\mu$, $v-s$, $v-p$ and $v-\mu$ are depicted by the coexistence areas. The excitation energy per nucleon in the caloric curve is discontinuous in the canonical ensemble at constant value of temperature and fixed values of the specific volume. The pressure in the isotherms $p-\rho$ and the chemical potential in the dependence $\mu-\rho$ are discontinuous at constat value of the specific volume and fixed values of the temperature.

We have shown that the first order phase transitions can be classified by the thermodynamic potentials of the statistical ensembles. We have also found that the first order phase transition associated with the Helmholtz free energy in the canonical ensemble requires the discontinuity of not only the extensive variables but also the intensive one. The Gibbs phase rule for this phase transition was formulated. It differs from the usual Gibbs phase rule of the first order phase transition of the liquid-gas type associated with the Gibbs free energy. The first order phase transition associated with the Helmholtz free energy is a consequence of thermal equilibrium between the system of nuclei and the environment which can exchange only heat with neither particle exchange nor work.

{\bf Acknowledgments:} This work was supported in part by the joint research project of JINR and IFIN-HH, protocol N~4063. I would like to acknowledge valuable remarks and fruitful discussions with A.S.~Sorin and D.V.~Anghel.

%\newpage

\end{document}